\documentclass[12pt]{iopart}
\usepackage{iopams}
\usepackage{graphicx} % for figures
\begin{document}

%-------------------------------------------------------------------------------
% Front Matter
%-------------------------------------------------------------------------------

\title{Nonlinear Quantum Search Using the Gross-Pitaevskii Equation}

\author{David A Meyer$^1$ and Thomas G Wong$^2$}
\address{$^1$ Department of Mathematics, University of California, San Diego, La Jolla, CA 92093-0112, USA}
\address{$^2$ Department of Physics, University of California, San Diego, La Jolla, CA 92093-0354, USA}
\ead{\mailto{dmeyer@math.ucsd.edu}, \mailto{tgw002@physics.ucsd.edu}}

\begin{abstract}
	We solve the unstructured search problem in constant time by computing with a physically motivated nonlinearity of the Gross-Pitaevskii type. This speedup comes, however, at the novel expense of increasing the time-measurement precision. Jointly optimizing these resource requirements results in an overall scaling of $N^{1/4}$. This is a significant, but not unreasonable, improvement over the $N^{1/2}$ scaling of Grover's algorithm. Since the Gross-Pitaevskii equation approximates the multi-particle (linear) Schr\"odinger equation, for which Grover's algorithm is optimal, our result leads to a quantum information-theoretic lower bound on the number of particles needed for this approximation to hold, asymptotically.
\end{abstract}


%\maketitle

%-------------------------------------------------------------------------------
% Main Matter
%-------------------------------------------------------------------------------

\section{Introduction}

Abrams and Lloyd \cite{AL1998} argued that a nonlinear quantum theory could result in unreasonable computational advantages by giving two examples of nonlinear algorithms that solve NP-complete and \#P problems in polynomial time. Both of their algorithms can be implemented by a nonlinear Schr\"odinger-type evolution in which the time derivatives of the state components depend upon their hyperbolic tangents \cite{C1998a, C1998b}. The derivative of $\tanh x$ at $x=0$ is $1$, so this is a strongly nonlinear system in which $0$ is an unstable fixed point. The strength of the nonlinearity provides a large computational advantage, but it also makes the system highly susceptible to noise \cite{AL1998,C1998a,C1998b}.

An obvious question is whether a modest, physically motivated nonlinearity can still produce a computational advantage. In particular, consider Bose-Einstein condensates (BECs). In 1924-25, Bose and Einstein predicted that cooling a dilute gas of bosons near absolute zero would cause the atoms to occupy their lowest quantum state, forming a new state of matter where quantum effects would be macroscopically apparent \cite{Bose1924, Einstein1924, Einstein1925}. It took seventy years, however, for these BECs to be experimentally produced \cite{Cornell1995, Ketterle1995, Hulet1995}. In general, describing such many-body systems is difficult because of the many interaction terms. But under certain conditions, one can assume that only two-body contact interactions contribute and the $s$-wave scattering length $a$ is much less than the interparticle spacing. Then using mean field theory, one finds that the system is approximately described by a nonlinear Schr\"odinger equation with a cubic nonlinearity:
\begin{equation}
	\label{eq:gpwave}
	\rmi \hbar \frac{\partial}{\partial t} \psi(\mathbf{r},t) = \left[ H_0 + \frac{4\pi\hbar^2a}{m} N_0 |\psi(\mathbf{r},t)|^2 \right] \psi(\mathbf{r},t),
\end{equation}
where $H_0$ includes the kinetic energy and trapping potential, $m$ is the mass of the condensate atom, and $N_0$ is the number of condensate atoms.\footnote{More generally, the cubic nonlinear Schr\"odinger equation is the equation of motion for $\phi^4$ field theory.} The validity of this celebrated Gross-Pitaevskii equation \cite{G1961, P1961} requires that $N_0$ be much greater than $1$---but how much greater?

We provide a quantum information-theoretic solution to this question in the context of solving the unstructured quantum search problem \cite{Grover1996} in continuous-time \cite{FG1998} using the Gross-Pitaevskii equation as the governing equation. The cubic nonlinearity in \eref{eq:gpwave} has zero derivative at zero, making it softer than those considered by Abrams and Lloyd \cite{AL1998}. We quantify the computational advantage that such a nonlinearity provides for the unstructured search problem compared to standard quantum computation; this requires considering time-measurement precision as a physical resource. Since this advantage cannot persist when the Gross-Pitaevskii equation is recognized as an approximation to an underlying multi-particle Schr\"odinger equation, for which Grover's algorithm is optimal, we arrive at a quantum information-theoretic lower bound on the number of particles, $N_0$, needed for this approximation to hold, asymptotically.

\section{Setup}

The system evolves in a $N$-dimensional Hilbert space with computational basis $\{ | 0 \rangle, \dots, | N-1 \rangle\}$. The initial state $| \psi(0) \rangle$ is an equal superposition $| s \rangle$ of all these basis states:
\[ | \psi(0) \rangle = | s \rangle = \frac{1}{\sqrt{N}} \sum_{i=0}^{N-1} | i \rangle. \]
The goal is to find a particular ``marked'' basis state, which we label $| w \rangle$.

\begin{figure}
	\begin{center}
		\includegraphics{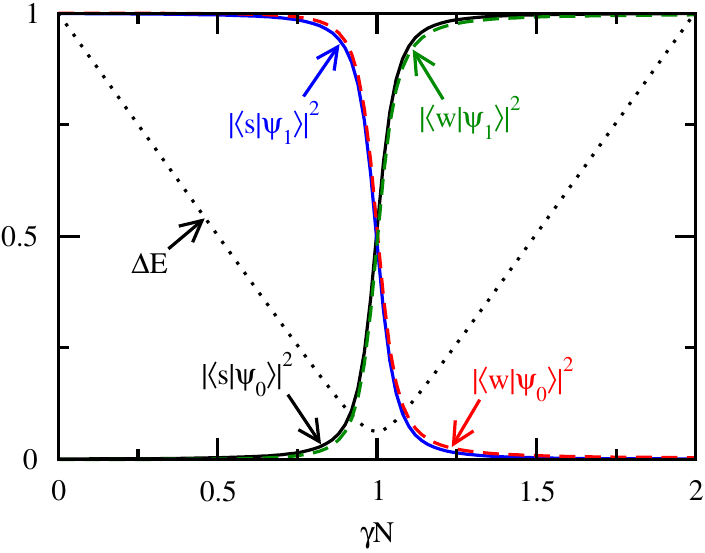}
		\caption{\label{fig:overlap_linear}Eigenvalue gap and eigenstate overlaps of $-H_0$ with $N = 1024$.}
	\end{center}
\end{figure}

Let's first review the linear solution. We use Childs and Goldstone's \cite{CG2004} notation for Farhi and Gutmann's \cite{FG1998} Hamiltonian:
\[ H_0 = -\gamma N | s \rangle \langle s | - | w \rangle \langle w |, \]
where $\gamma$ is a parameter, inversely proportional to mass. The system evolves in the two-dimensional subspace spanned by $| w \rangle$ and $| s \rangle$. One might (correctly) reason that the success of the algorithm depends on the value of $\gamma$. This can be seen in \fref{fig:overlap_linear}, which shows the difference in eigenvalues of $-H_0$ and the overlaps of its nontrivial eigenvectors with $| s \rangle$ and $| w \rangle$. When $\gamma$ takes a critical value of $\gamma_{\rm c} = 1/N$, the eigenstates of $-H_0$ are proportional to $\pm | w \rangle + | s \rangle$, and the corresponding energy gap is $2/\sqrt{N}$. So the Schr\"odinger evolution rotates the state from $| s \rangle$ to $| w \rangle$ in time $\pi \sqrt{N} / 2$.

In the nonlinear regime, we include an additional nonlinear ``self-potential'' $V(t)$ so that the system evolves according to the Gross-Pitaevskii equation \eref{eq:gpwave}:
\[ \rmi \frac{\partial}{\partial t} \psi(\mathbf{r},t) = \big[ H_0 - \underbrace{g |\psi(\mathbf{r},t)|^2}_{V(t)} \big] \psi(\mathbf{r},t), \]
where $g > 0$. This corresponds to a BEC with attractive interactions, and thus a negative scattering length \cite{Stoof1994, Bradley1995}. Heuristically, as probability accumulates at the marked state due to the $| w \rangle \langle w |$ term in $H_0$, the self-potential attracts more probability, speeding up the search. Thus we expect larger $g$ to result in a faster algorithm.

In the computational basis, the self-potential is
\[ V(t) = g \sum_{i=0}^{N-1} \left| \langle i | \psi \rangle \right|^2 | i \rangle \langle i |. \]
Even with this nonlinearity, the system remains in the subspace spanned by $\{ | w \rangle, | s \rangle \}$ throughout its evolution. We define a vector
\[ | r \rangle = \frac{1}{\sqrt{N-1}} \sum_{i \ne w} | i \rangle, \]
which is orthonormal to $| w \rangle$. Then the state of the system $| \psi(t) \rangle$ can be written as
\[ | \psi(t) \rangle = \alpha(t) | w \rangle + \beta(t) | r \rangle. \]
Writing the Gross-Pitaevskii equation in this $\{ | w \rangle, | r \rangle\}$ basis, we get
\begin{eqnarray}
	\frac{\rmd}{\rmd t} \left( \begin{array}{c} \alpha \\ \beta \end{array} \right)
		&= -\rmi \left( H_0 - V \right) \left( \begin{array}{c} \alpha \\ \beta \end{array} \right) \label{eq:gp}  \nonumber \\
		&= \rmi \underbrace{ \left( \begin{array}{cc}
			\gamma + 1 + g|\alpha|^2 & \gamma \sqrt{N-1} \\
			\gamma \sqrt{N-1} & \gamma (N-1) + \frac{g}{N-1} |\beta|^2
		  \end{array} \right)}_A
		  \left( \begin{array}{c} \alpha \\ \beta \end{array} \right),
\end{eqnarray}
where we've defined $A = -(H_0 - V)$.

\section{Critical Gamma}

\begin{figure}
	\begin{center}
		\includegraphics{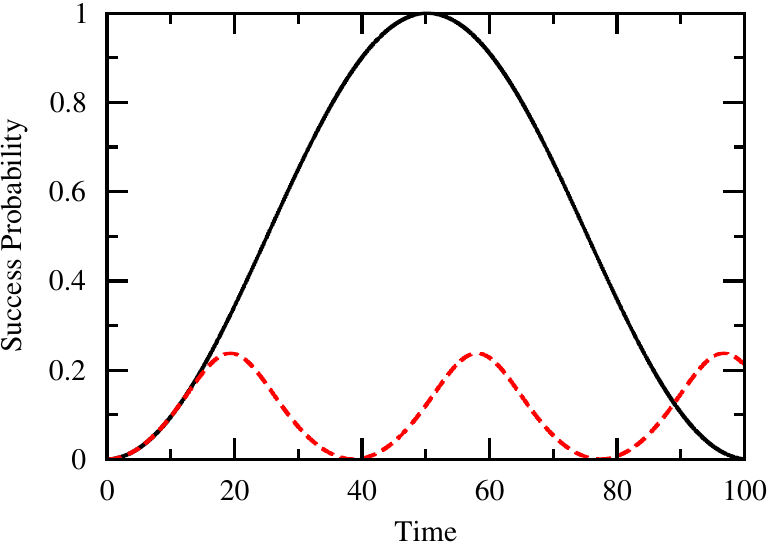}
		\caption{\label{fig:prob_time}Success probability as a function of time for $N = 1024$ and $\gamma = 1/N$ constant. The solid curve is the linear ($g = 0$) case, and the dashed curve is the nonlinear $g = 1$ case.}
	\end{center}
\end{figure}

\begin{figure}
	\begin{center}
		\includegraphics{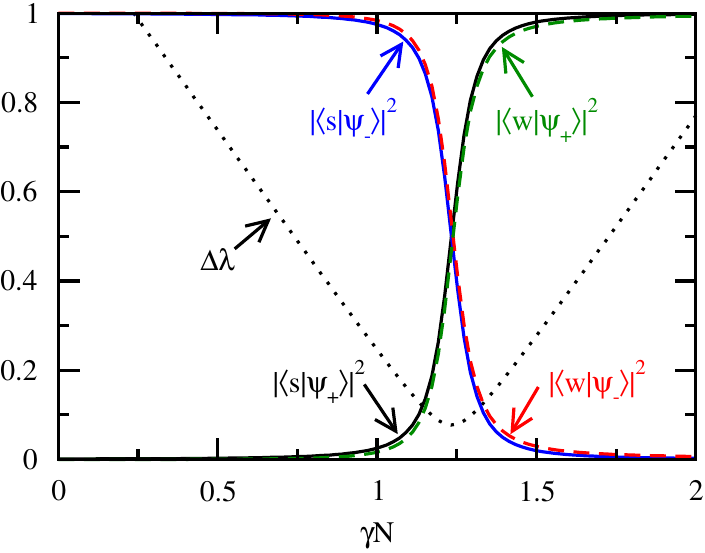}
		\caption{\label{fig:overlap_nonlinear}Eigenvalue gap and eigenstate overlaps of $A$ at $t = 20$ for nonlinear search with $N = 1024$, $g = 1$, and $\gamma = 1/N$ constant.}
	\end{center}
\end{figure}

Before proceeding with further analytical calculations, we build some intuition by examining two plots. For constant $\gamma$ and $g$, the success probability as a function of time, $|\alpha(t)|^2$, is plotted in \fref{fig:prob_time} along with the linear result. The nonlinear algorithm underperforms the linear one in this case. This is true in general for constant $\gamma$ and $g$, and it can be understood by examining the time-dependence of the critical value of $\gamma$, which is the value of $\gamma$ that ensures that the eigenstates of $A$ are in the form $\pm | w \rangle + | s \rangle$. Initially, $\gamma_{\rm c} = 1/N$. Then, as shown in \fref{fig:overlap_nonlinear}, it shifts to a larger value. If $\gamma$ is constant, it will not follow this shift, we will no longer have the desired eigenstates, and the algorithm will perform poorly.

To determine how $\gamma_{\rm c}$ varies with time, we find the eigenvectors of $A$ and choose $\gamma$ so that they have the desired form $\pm | w \rangle + | s \rangle$. To eliminate fractions in the subsequent algebra, we rescale the nonlinearity coefficient $g$ by defining
\[ G = \frac{g}{N-1}. \]
Solving the characteristic equation gives the eigenvalues of $A$:
\[ \lambda_\pm = \frac{1}{2} \left( \gamma N + 1 + G \sigma \right) \pm \frac{1}{2} \Delta\lambda, \]
where the gap between them is
\[ \Delta\lambda = \sqrt{(\gamma N - 1)^2 + 4\gamma + G^2 \delta^2 + 2G\delta \left[ 1 - \gamma(N-2) \right]}, \]
and we've defined
\[ \sigma = (N-1)|\alpha|^2 + |\beta|^2 \quad \mathrm{and} \quad \delta = (N-1)|\alpha|^2 - |\beta|^2. \]
The corresponding eigenvectors of $A$ are
\[ | \psi_\pm \rangle = \sqrt{\frac{N}{N-1}} \left[ \frac{-\gamma N + 1 + \delta G \pm \Delta\lambda}{2 \gamma \sqrt{N}} | w \rangle + | s \rangle \right]. \]
The critical value of $\gamma$ ensures that these eigenvectors have the form $\pm | w \rangle + | s \rangle$. That is,
\[ \left. \frac{-\gamma N + 1 + \delta G \pm \Delta\lambda}{2 \gamma \sqrt{N}} \right|_{\gamma_{\rm c}} = 1. \]
Solving this yields:
\begin{equation}
	\label{eq:criticalgamma}
	\gamma_{\rm c} = \frac{1 + G \delta}{N}.
\end{equation}
Note that in the linear limit ($G=0$), this reduces to $\gamma_{\rm c} = 1/N$, as expected. Importantly, since $\delta$ varies with time, \eref{eq:criticalgamma} implies $\gamma_{\rm c}$ also varies with time, in agreement with our previous discussion about figures \ref{fig:prob_time} and \ref{fig:overlap_nonlinear}.

\section{Runtime}

\begin{figure}
	\begin{center}
		\includegraphics{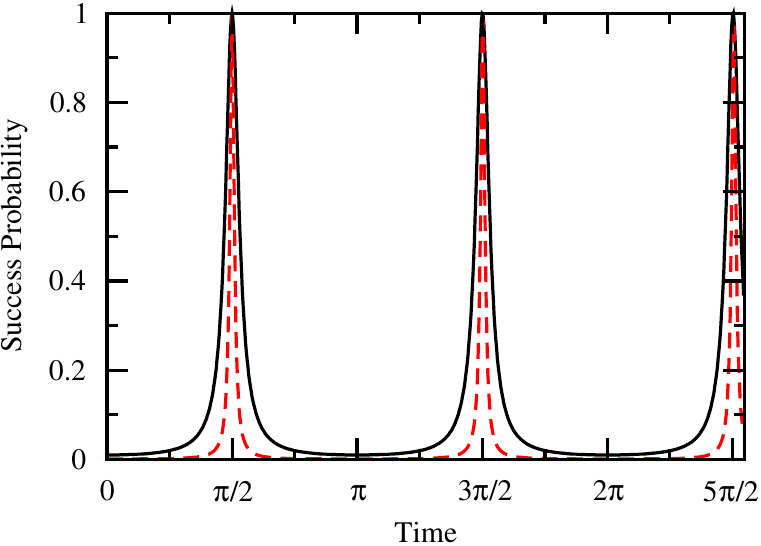}
		\caption{\label{fig:prob_time_critical}Success probability as a function of time for nonlinear search with $G = 1$ and $\gamma = \gamma_{\rm c}$ as defined in \eref{eq:criticalgamma}. The solid line is $N = 100$ and the dashed line is $N = 1000$.}
	\end{center}
\end{figure}

For the remainder of the paper, we choose time-varying $\gamma = \gamma_{\rm c}$ according to \eref{eq:criticalgamma}. Before analytically determining the consequences of this, let's again consider a plot. \Fref{fig:prob_time_critical} shows the success probability as a function of time. There are several observations. First, the success probability reaches $1$, which occurs because we constructed the eigenstates to make this happen. Second, as $N$ increases, the runtime remains constant. Third, the success probability is periodic. Finally, the peak in success probability becomes increasingly narrow for large $N$. Let's now analytically prove the second, third, and fourth observations.

To begin, we explicitly write out \eref{eq:gp} to get two coupled, first-order ordinary differential equations for $\alpha(t)$ and $\beta(t)$:
\begin{eqnarray}
	\label{eq:dadt} \frac{\rmd \alpha}{\rmd t} = \rmi \left\{ \left[ \gamma_{\rm c} + 1 + G(N-1) |\alpha|^2 \right] \alpha + \gamma_{\rm c} \sqrt{N-1} \beta \right\} \\
	\label{eq:dbdt} \frac{\rmd \beta}{\rmd t} = \rmi \left\{ \gamma_{\rm c} \sqrt{N-1} \alpha + \left[ \gamma_{\rm c} (N-1) + G |\beta|^2 \right] \beta \right\}.
\end{eqnarray}
We can decouple these equations for the success probability $x(t) = |\alpha(t)|^2$, yielding
\begin{equation}
	\label{eq:firstderiv}
	\frac{\rmd x}{\rmd t} = \pm \sqrt{ \frac{4(Nx-1)(1-x)\left[1+G(Nx-1)\right]^2}{N^2} }.
\end{equation}
The details of this decoupling procedure are given in \ref{appendix}. To solve this uncoupled equation, we use separation of variables and integrate from $t = 0$ to $t$ and $x = 1/N$ to $x$, which yields
\begin{equation}
	\label{eq:timeprob}
	t = -\sqrt{\frac{N}{1+G(N-1)}} \left\{ \tan^{-1}\left[\frac{\sqrt{N} \sqrt{1-x}}{\sqrt{1+G(N-1)} \sqrt{Nx-1}}\right] - \frac{\pi}{2} \right\}.
\end{equation}
Solving for $x$, the success probability as a function of time is
\begin{equation}
	\label{eq:probtime}
	x(t) = \frac{N + \left[ 1+G(N-1) \right] \tan^2 \left[ \frac{\pi}{2} - \sqrt{\frac{1+G(N-1)}{N}} t \right]}{N + N  \left[ 1+G(N-1) \right] \tan^2 \left[ \frac{\pi}{2} - \sqrt{\frac{1+G(N-1)}{N}} t \right]} .
\end{equation}
From this, the success probability reaches $1$ when the tangent term is zero, which first occurs at time
\[ t_* = \frac{1}{\sqrt{1+G (N-1)}} \frac{\pi \sqrt{N}}{2}. \]
This runtime is exactly constant for $G = 1$. Also, when $G = \Theta(1)$, the runtime for large $N$ is $\pi/2\sqrt{G}$, and thus asymptotically constant (and arbitrarily small!). From \eref{eq:probtime}, we also see that the success probability is periodic with a period of $2t_*$.

Now let's prove that the peak in success probability is narrow by finding its width, thus proving all our observations about \fref{fig:prob_time_critical}. Using \eref{eq:timeprob}, the difference in time at which the success probability reaches a height of $1 - \epsilon$ is
\[ \Delta t = 2 \sqrt{\frac{N}{1+G(N-1) }} \tan^{-1}\left[\frac{\sqrt{N} \sqrt{\epsilon}}{\sqrt{1+G(N-1)} \sqrt{N(1-\epsilon)-1}}\right]. \]
The $\tan^{-1}$ makes it difficult to determine the scaling with $N$, so we Taylor expand it:
\[ \Delta t = \frac{2N}{1+G(N-1)} \sqrt{\frac{\epsilon}{N-1}} + O(\epsilon^{3/2}). \]
When $G = N^\kappa$, the first term scales as $\Theta(N^{1/2})$ when $\kappa \le -1$ and $\Theta(N^{-1/2 - \kappa})$ when $\kappa > -1$, for large $N$. To determine whether keeping this first term alone is sufficient, we use Taylor's remainder theorem to bound the error
\[ R_1(\epsilon) \le \frac{N^2 (1+3G(N(1-\epsilon)-1))}{\left(N (1-\epsilon)-1\right)^{3/2} \left(1+G\left(N \left(1-\epsilon\right)-1\right)\right)^2} \epsilon^{3/2}, \]
which has the same scaling for large $N$ as the first term in the Taylor series for $\Delta t$. Thus it suffices to keep only the first term.

For constant $G$, the width in success probability is $\Theta(1/\sqrt{N})$, which agrees with our observation from \fref{fig:prob_time_critical} that the peak in success probability is increasingly narrow as $N$ increases. Thus we must measure the system with increasing time precision. This behavior is opposite the linear case. That is, when $G = 0$ the width is $\Theta(\sqrt{N})$, so the time at which we measure the result can be increasingly imprecise as $N$ increases.

\section{Time-Measurement Precision}

This time-measurement precision requirement of the nonlinear algorithm requires additional resources. In particular, time and frequency standards are currently defined by atomic clocks, such as NIST-F1 in the United States \cite{NIST-F1}. An atomic clock with $n_\mathrm{clock}$ ions used as atomic oscillators has a time-measurement precision of $1/\sqrt{n_\mathrm{clock}}$ when the ions are acted upon independently. This can be improved using quantum entanglement, reducing the time-measurement precision to $1/n_\mathrm{clock}$ \cite{BIWH1996, GLM2004}. Even with this improvement, our constant-time nonlinear search algorithm would require $O(\sqrt{N})$ ions in an atomic clock to have sufficiently high time-measurement precision to measure the peak in success probability. So, although our nonlinear algorithm runs in constant time, the total resource requirement is still $O(\sqrt{N})$, the same as the linear algorithm. This raises the possibility that nonlinear quantum mechanics may not provide efficient solutions to NP-complete and \#P problems when all the resource requirements are taken into consideration \cite{AL1998}.

\begin{figure}
	\begin{center}
		\includegraphics{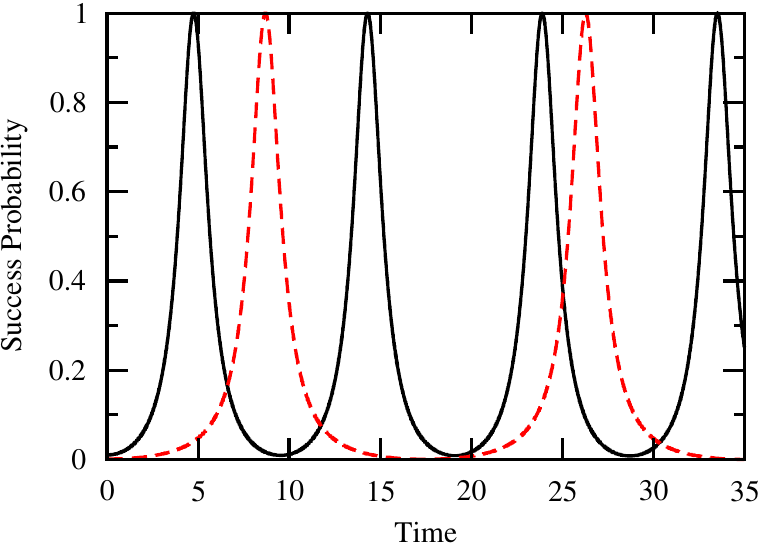}
		\caption{\label{fig:prob_time_jointoptimized}Success probability as a function of time for nonlinear search with $G = N^{-1/2}$ and $\gamma = \gamma_{\rm c}$ as defined in \eref{eq:criticalgamma}. The solid line is $N = 100$ and the dashed line is $N = 1000$. The peaks have same width, independent of $N$.}
	\end{center}
\end{figure}

In our case, however, we can settle for a smaller improvement in runtime and reduce the time-measurement precision and total resource requirement. If we let $G$ decrease as $N^{\kappa}$ for $\kappa\le0$, then the runtime is $t_* = \Theta(N^{-\kappa/2})$, and the time-measurement precision is $\Delta t = \Theta(N^{-1/2-\kappa})$, where we've assumed for both that $\kappa > -1$, since for $\kappa\le -1$, $\Delta t = \Theta(N^{1/2})$, independently of $G$. This time-measurement precision requires $O(N^{1/2+\kappa})$ ions in an atomic clock that utilizes entanglement. We assume, as in the setup for Grover's algorithm, that $\log N$ qubits can be used to encode the $N$-dimensional Hilbert space; these should also be included in the required ``space'' resources. Multiplying the time and ``space'' requirements together, which preserves the time-space tradeoff inherent in na\"ive parallelization, the resulting total resource requirement takes a minimum value of $O(N^{1/4} \log N)$ when $\kappa = -1/2$ (so that the runtime is $N^{1/4}$ and the time-measurement precision is constant). The success probability as a function of time at this jointly optimized value of $G$ is plotted in \fref{fig:prob_time_jointoptimized}; note that the peak width is independent of $N$.

This significant---but not unreasonable---improvement over the $\Theta(\sqrt{N} \log N)$ time-space resource requirements of the linear quantum search algorithm is consistent with our expectation that a modest nonlinearity should result in a modest speedup.

\section{Repulsive Interactions}

Our nonlinear search algorithm was based on the intuition that attractive interactions speed up the accumulation of success probability. By the same intuition, repulsive interactions, where $G < 0$, should yield a worse runtime. Our derivation of \eref{eq:criticalgamma} for the critical value of $\gamma$ is unchanged if we flip the sign of $G$, so \eref{eq:uncoupled} and \eref{eq:firstderiv} are still valid for repulsive interactions. These equations yield critical points $x_* = 1/N$, $1$, and $(G-1)/NG$, corresponding to minima, maxima, and stationary points, respectively.

\begin{figure}
	\begin{center}
		\includegraphics{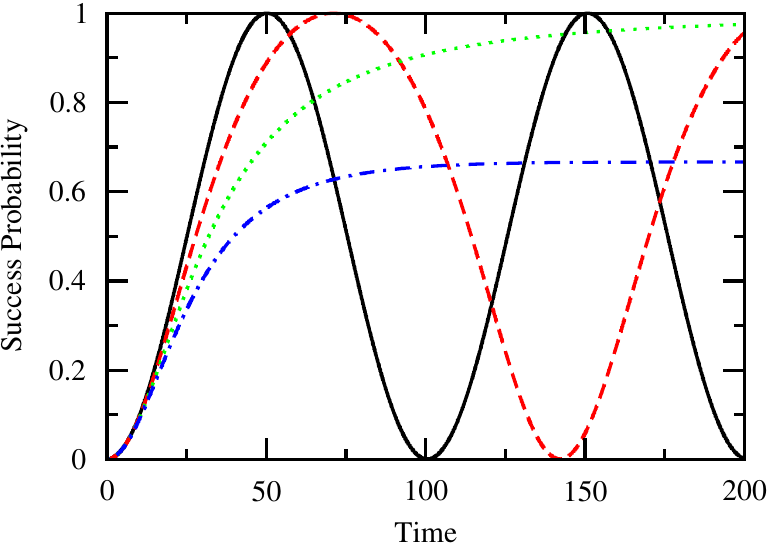}
		\caption{\label{fig:prob_time_repulsive}Success probability as a function of time for $N = 1024$ and $\gamma = \gamma_{\rm c}$ as defined in \eref{eq:criticalgamma}. The solid line is the linear ($g=0$) case, the dashed line is the nonlinear $g = -0.5$ case, the dotted line is the nonlinear $g = -1$ case, and the dot-dashed line is the nonlinear $g = -1.5$ case.}
	\end{center}
\end{figure}

When $G > -1/(N-1)$, the success probability is unhindered by the stationary point and reaches a maximum value of $1$, as shown in the dashed curve of \fref{fig:prob_time_repulsive}. When $G < -1/(N-1)$, however, reaching this maximum is precluded by the presence of a stationary point, as shown in the dashed and dot-dashed curves of \fref{fig:prob_time_repulsive}.

We can explicitly prove that repulsive interations ($G < 0$) will underperform the linear ($G=0$) algorithm. From \eref{eq:firstderiv},
\[ \frac{\rmd x}{\rmd t} = \pm \frac{2}{N} \sqrt{(Nx-1)(1-x)} \left[ 1 + G(Nx-1) \right]. \]
So when $G < 0$, the magnitude of $\rmd x/\rmd t$ at a particular value of $x$ is less than when $G = 0$. Then success probability will increase more slowly for repulsive interactions (except initially, where they increase at the same rate). Thus it will underperform the linear algorithm.

\section{Validity of the Gross-Pitaevskii Equation}

Of course, the cubic nonlinearity we've exploited is not fundamental, but rather occurs in an effective description of an interacting multi-particle quantum system (\textit{e.g.}, a BEC). So we must include the number of particles $N_0$ in our resource accounting. Each particle interacts with the potential at the marked site, so in the framework of Zalka's optimality proof for Grover's algorithm \cite{Zalka1999} (generalized to continuous time \cite{Cleve2009}), there are $N_0$ oracles, each responding to a $\log N$ bit query.  Zalka showed that the product of the space requirements and the square of the time requirements is lower bounded by $N$, \textit{i.e.}, $(N_0 \log N)(N^{1/4})^2 = \Omega(N)$. Solving for the number of particles, $N_0 = \Omega(N^{1/2}/\log N)$. This is a quantum information-theoretic lower bound on the number of particles necessary for the Gross-Pitaevskii equation to be the correct asymptotic description of the multi-particle (linear) quantum dynamics.

Notice that once we account for the scaling of $N_0$ in the space requirements, the product of the time and space requirements is $O(N^{3/4})$, worse than the $O(N^{1/2}\log N)$ of Grover's algorithm.  In fact, if we calculate for the general case $G = N^{\kappa}$, where $\kappa$ need not be chosen to optimize the product of the time and space (ignoring $N_0$) resources, Zalka's bound implies $N_0 = \Omega(\max\{1,N^{1+\kappa}/\log N\})$, so the total time-space requirements are $O(N^{1+\kappa/2})$ for $\kappa> -1$, and $O(N^{1/2}\log N)$ when $\kappa = -1$.  This is optimized for $\kappa = -1$, {\it i.e.}, by Grover's algorithm.  On the other hand, Zalka's bound is strongest when $\kappa = 0$, in which case it implies that $N_0 = \Omega(N/\log N)$. That is, the existence of the constant time nonlinear algorithm we found in section 4 implies this stronger lower bound on $N_0$, despite the $O(N^{1/2})$ number of clock ions required.  To our knowledge, this is the first lower bound derived on the scaling of $N_0$ required for the Gross-Pitaevskii equation be a good asymptotic approximation.

This bound also is significantly stronger than the bound implied by the physically plausible requirement that the volume of the multi-particle condensate, and thus $N_0$, be of at least the order of the volume of space in which the $N$ possible discrete locations are defined.  Were we working in any fixed, finite dimension, {\it e.g.}, on a cubic lattice, the volume would be proportional to $N$, implying $N_0 = \Omega(N)$.  But we are not; the complete graph with equal pairwise transition rates is realized by the vertices and edges of an equilateral $(N-1)$-dimensional simplex.  With edges of length 1, this has volume $\sqrt{N/2^{N-1}}/(N-1)!$, which is much smaller than $N$, and also much smaller than our bound of $N/\log N$.

\section{Critical Gamma is a Continuous Rescaling of Time}

We previously derived the critical value of $\gamma$ so that the eigenstates of the Hamiltonian are proportional to $\pm | w \rangle + | s \rangle$. Now we examine what the critical value of $\gamma$ does from another perspective. Recall the ``Hamiltonian'' we've been using is
\[ H = - \gamma N | s \rangle \langle s | - | w \rangle \langle w | - g \sum_i |\psi_i|^2 | i \rangle \langle i |, \]
where $\psi_i = \langle i | \psi \rangle$. Explicitly writing the nonlinear term as marked and unmarked terms, we get
\begin{eqnarray*}
	H
		&= - \gamma N | s \rangle \langle s | - | w \rangle \langle w | - g |\alpha|^2 | w \rangle \langle w | - g \frac{|\beta|^2}{N-1} \sum_{x \ne w} | x \rangle \langle x | \\
		&= - \gamma N | s \rangle \langle s | - | w \rangle \langle w | - G(N-1) |\alpha|^2 | w \rangle \langle w | - G |\beta|^2 \sum_{x \ne w} | x \rangle \langle x | \\
		&= - \gamma N | s \rangle \langle s | - \left[ 1 + G(N-1) |\alpha|^2 \right] | w \rangle \langle w | - G |\beta|^2 \sum_{x \ne w} | x \rangle \langle x |.
\end{eqnarray*}
Recall $\gamma = \gamma_{\rm c}$ is chosen according to \eref{eq:criticalgamma}:
\[ \gamma_{\rm c} N = 1 + G(N-1)|\alpha|^2 - G|\beta|^2, \]
which we arrange to get
\[ 1 + G(N-1)|\alpha|^2 = \gamma_{\rm c} N + G|\beta|^2. \]
Then the Hamiltonian becomes
\begin{eqnarray*}
	H
		&= - \gamma_{\rm c} N | s \rangle \langle s | - \left[ \gamma_{\rm c} N + G |\beta|^2 \right] | w \rangle \langle w | - G |\beta|^2 \sum_{x \ne w} | x \rangle \langle x | \\
		&= - \gamma_{\rm c} N \left( | s \rangle \langle s | + | w \rangle \langle w | \right) - G |\beta|^2 \mathbb{I} .
\end{eqnarray*}
The last term continuously redefines the ``zero'' of energy, so we can drop it. That is, it only changes the overall phase of the system, which has no measurable effect. Then the Hamiltonian is
\[ H = - \gamma N \left( | s \rangle \langle s | + | w \rangle \langle w | \right). \] 
Importantly, $H_{\rm FG} = -| s \rangle \langle s | - | w \rangle \langle w |$ is the Hamiltonian from Farhi and Gutmann's ``analog analogue'' of Grover's algorithm \cite{FG1998}, and it is optimal. Our nonlinear algorithm has a factor of $\gamma N$, so it effectively follows their optimal algorithm, but with a continuously rescaled time. That is, the system evolves according to
\[ \rmi \frac{\rmd\psi}{\gamma N \rmd t} = H_{\rm FG} \psi. \]
Let's call the rescaled time $\tau(t)$ so that $\rmd\tau = \gamma N \rmd t$. Then
\[ \tau = \int \! \gamma N \rmd t, \]
and the equation of motion becomes
\[ i \frac{\rmd\psi}{\rmd\tau} = H_{\rm FG} \psi. \]
This has success probability given by (11) of \cite{FG1998}:
\[ x(\tau) = \sin^2 \left( \frac{\tau}{\sqrt{N}} \right) + \frac{1}{N} \cos^2 \left( \frac{\tau}{\sqrt{N}} \right). \]
Plugging in for $\tau$,
\[ x(t) = \sin^2 \left( \frac{\int \! \gamma N \rmd t}{\sqrt{N}} \right) + \frac{1}{N} \cos^2 \left( \frac{\int \! \gamma N \rmd t}{\sqrt{N}} \right). \]
Since $\gamma_{\rm c} N = 1 + G\delta = 1 - G + GNx$, we get
\[ x(t) = \sin^2 \left( \frac{(1-G)t + GN \int \! x(t) \rmd t}{\sqrt{N}} \right) + \frac{1}{N} \cos^2 \left( \frac{(1-G)t + GN \int \! x(t) \rmd t}{\sqrt{N}} \right). \]
This integral transcendental equation gives $x(t)$. While the difficulty of solving this equation makes it less useful in practice, it does reveal our nonlinear algorithm's relationship with the linear, optimal algorithm. In particular, a different control policy for $\gamma$ will cause the system to evolve along a different, slower path. While not a proof, this is an argument for the optimality of our algorithm.

\section{Multiple Marked States}

Our analysis naturally extends to the case of $k$ marked states. Let $M$ be the set of marked basis states. As before, the system evolves in a two-dimensional subspace:
\[ | \psi(t) \rangle = \alpha(t) \frac{1}{\sqrt{k}} \sum_{x \in M} | x \rangle + \beta(t) \frac{1}{\sqrt{N-k}} \sum_{x \notin M} | x \rangle. \]
The system evolves according to
\[ \frac{\rmd}{\rmd t} | \psi(t) \rangle = \rmi A | \psi \rangle, \]
where
\[ A = \gamma N | s \rangle \langle s | + \left( 1+g\frac{|\alpha|^2}{k} \right) \sum_{x \in M} | x \rangle \langle x | + g \frac{|\beta|^2}{N-k} \sum_{x \notin M} | x \rangle \langle x | \]
includes both the linear Hamiltonian and the nonlinear ``self-potential''. The eigenstates of $A$ have the form $\pm | w \rangle + | s \rangle$ when $\gamma$ is
\[ \gamma_{\rm c} = \frac{1 + G \delta}{N}, \]
where $G = g/(k(N-k))$ and $\delta = (N-k) |\alpha|^2 - k |\beta|^2$.
At $\gamma = \gamma_{\rm c}$, we can decouple these equations in the same manner as the $k = 1$ case (see \ref{appendix}) and integrate from $t = 0$ to $t$ and $x = k/N$ to $x$ to get
\[ t = -\sqrt{\frac{N}{k (1+G (N-k)) }} \left\{ \tan^{-1}\left[\frac{\sqrt{N} \sqrt{1-x}}{\sqrt{1+G(N-k)} \sqrt{N x-k}}\right] - \frac{\pi}{2} \right\}, \]
which can be solved for a success probability of
\[ x(t) = \frac{N + k \left[ 1+G(N-k) \right] \tan^2 \left[ \frac{\pi}{2} - \sqrt{\frac{k(1+G(N-k))}{N}} t \right]}{N + N  \left[ 1+G(N-k) \right] \tan^2 \left[ \frac{\pi}{2} - \sqrt{\frac{k(1+G(N-k))}{N}} t \right]}. \]
Then the runtime is
\[ t_* = \frac{1}{\sqrt{k(1+G (N-k))}} \frac{\pi \sqrt{N}}{2}, \]
and the success probability is still periodic with period $2t_*$. At this runtime, the peak in success probability has a width of
\[ \Delta t = 2 \sqrt{\frac{N}{k (1+G (N-k)) }} \tan^{-1}\left[\frac{\sqrt{N} \sqrt{\epsilon}}{\sqrt{1+G(N-k)} \sqrt{N(1-\epsilon)-k}}\right], \]
but Taylor's theorem can be used to show that it suffices to keep the first term in the Taylor series:
\[ \Delta t = \frac{2N}{1+G(N-k)} \sqrt{\frac{\epsilon}{k(N-k)}} + O(\epsilon^{3/2}). \]

As in the case of a single marked state, we can find the scaling of $G = N^\kappa$ that optimizes the product of ``space'' and time, where ``space'' includes both the number of ions needed in an atomic clock that utilizes entanglement to achive sufficiently high time-measurement precision, and the $\log N$ qubits needed to encode the $N$-dimensional Hilbert space. Say the number of marked sites scales as $k = N^\lambda$, with $0 \le \lambda \le 1$. When $\kappa = -\lambda/2 - 1/2$, the product of ``space'' and time takes a minimum value of $ST = N^{-\lambda/4+1/4} \log N$ (so that the runtime is $N^{-\lambda/4+1/4}$ and the time-measurement precision is constant). Note this is a square root speedup over the linear ($G = 0$) algorithm, whose product of ``space'' and time is $N^{-\lambda/2+1/2} \log N$. Thus our nonlinear method, by varying $\gamma$ and choosing an optimal nonlinear coefficient $G$, provides a significant, but not unreasonable, improvement over the continuous-time analogue of Grover's algorithm, even with multiple marked items.

%-------------------------------------------------------------------------------
% Acknowledgments
%-------------------------------------------------------------------------------

\ack
This work was partially supported by the Defense Advanced Research Projects Agency as part of the Quantum Entanglement Science and Technology program under grant N66001-09-1-2025, and by the Air Force Office of Scientific Research as part of the Transformational Computing in Aerospace Science and Engineering Initiative under grant FA9550-12-1-0046.

%-------------------------------------------------------------------------------
% Appendix
%-------------------------------------------------------------------------------

\appendix

\section{Decoupling the Equations of Motion}
\label{appendix}

We begin with two coupled, first-order ordinary differential equations 
\begin{eqnarray}
	\label{eq:dadt} \frac{\rmd \alpha}{\rmd t} = \rmi \left\{ \left[ \gamma_{\rm c} + 1 + G(N-1) |\alpha|^2 \right] \alpha + \gamma_{\rm c} \sqrt{N-1} \beta \right\} \\
	\label{eq:dbdt} \frac{\rmd \beta}{\rmd t} = \rmi \left\{ \gamma_{\rm c} \sqrt{N-1} \alpha + \left[ \gamma_{\rm c} (N-1) + G |\beta|^2 \right] \beta \right\}.
\end{eqnarray}
We decouple these equations by defining three real variables $x(t)$, $y(t)$, and $z(t)$ such that
\begin{eqnarray}
	\label{eq:x} x = |\alpha|^2 \\
	\label{eq:yz} y + \rmi z = \alpha \beta^*.
\end{eqnarray}
Note that $x(t)$ defined by \eref{eq:x} is the success probability. Differentiating it and utilizing \eref{eq:dadt}, we find that
\[ \frac{\rmd x}{\rmd t} = \frac{\rmd |\alpha|^2}{\rmd t} = \alpha \frac{\rmd \alpha^*}{\rmd t} + \frac{\rmd \alpha}{\rmd t} \alpha^* = 2 \gamma_{\rm c} \sqrt{N-1} z. \]
Solving this for $z$, we get
\begin{equation}
	\label{eq:z}
	z = \frac{1}{2 \gamma_{\rm c} \sqrt{N-1}} \frac{\rmd x}{\rmd t}.
\end{equation}
Noting that $\rmd \gamma_{\rm c}/\rmd t = G \, \rmd x/\rmd t$, we differentiate \eref{eq:z} to get
\begin{equation}
	\label{eq:dzdt1}
	\frac{\rmd z}{\rmd t} = \frac{1}{2 \sqrt{N-1}} \left[ \frac{-1}{\gamma_{\rm c}^2} G \left( \frac{\rmd x}{\rmd t} \right)^2 + \frac{1}{\gamma_{\rm c}} \frac{\rmd^2x}{\rmd t^2} \right].
\end{equation}
Now we want to find another expression for $\rmd z/\rmd t$, which we can then set equal to \eref{eq:dzdt1}. We do this by differentiating \eref{eq:yz}, utilizing \eref{eq:dadt} and \eref{eq:dbdt}, and equating the real and imaginary parts, which yields
\begin{eqnarray}
	\label{eq:dydt} \frac{\rmd y}{\rmd t} = - 2\gamma_{\rm c} z \\
	\label{eq:dzdt2} \frac{\rmd z}{\rmd t} = 2\gamma_{\rm c} y + \gamma_{\rm c} \sqrt{N-1} (1-2x).
\end{eqnarray}
Substituting \eref{eq:z} for $z$ into \eref{eq:dydt}, we get
\[ \frac{\rmd y}{\rmd t} = \frac{-1}{\sqrt{N-1}} \frac{\rmd x}{\rmd t}, \]
which integrates to
\[ y = \frac{1}{\sqrt{N-1}} (1-x), \]
where the constant of integration was found using $x(0) = 1/N$ and $y(0) = \sqrt{N-1}/N$. Now we can plug this into \eref{eq:dzdt2} to get
\begin{eqnarray*}
	\frac{\rmd z}{\rmd t} 
		&= 2 \gamma_{\rm c} \frac{1}{\sqrt{N-1}} (1-x) + \gamma_{\rm c} \sqrt{N-1} (1-2x) \\
		&= \frac{\gamma_{\rm c}}{\sqrt{N-1}} \left( 1 + N - 2Nx \right).
\end{eqnarray*}
Equating this to \eref{eq:dzdt1} and simplifying yields
\[ \frac{\rmd^2x}{\rmd t^2} = \frac{G}{\gamma_{\rm c}} \left( \frac{\rmd x}{\rmd t} \right)^2 + 2 \gamma_{\rm c}^2 \left( 1 + N - 2Nx \right). \]
Plugging in for $\gamma_{\rm c}$ as defined in \eref{eq:criticalgamma}, this becomes
\begin{equation}
	\label{eq:uncoupled}
	\frac{\rmd^2x}{\rmd t^2} = \frac{NG}{1-G+NGx} \left( \frac{\rmd x}{\rmd t} \right)^2 + \frac{2}{N^2} \left( 1-G+NGx \right)^2 \left( 1+N-2Nx \right)
\end{equation}
Now let $f(x) = (\rmd x/\rmd t)^2$ so that $\rmd f/\rmd x = 2 \rmd^2x/\rmd t^2$. Then \eref{eq:uncoupled} becomes
\[
	\frac{1}{2} \frac{\rmd f}{\rmd x} = \frac{NG}{1-G+NGx} f + \frac{2}{N^2} \left( 1-G+NGx \right)^2 \left( 1+N-2Nx \right).
\]
Solving this first-order ODE and using the initial condition $f(x=1/N) = 0$, we get
\[ f(x) = \frac{4(Nx-1)(1-x)\left[1+G(Nx-1)\right]^2}{N^2}. \]
Taking the square root and noting that $\rmd x/\rmd t = \pm \sqrt{f(x)}$,
\begin{equation}
	\frac{\rmd x}{\rmd t} = \pm \sqrt{ \frac{4(Nx-1)(1-x)\left[1+G(Nx-1)\right]^2}{N^2} }.
\end{equation}

%-------------------------------------------------------------------------------
% References.
%-------------------------------------------------------------------------------

\section*{References}
\bibliographystyle{iopart-num}
\bibliography{refs}

\end{document}